\documentclass[preprint,12pt]{elsarticle}

\usepackage[figuresright]{rotating}

\usepackage[T1]{fontenc}
\usepackage{graphicx}
\usepackage{todonotes}
\usepackage{amssymb}
\usepackage{amsmath}
\usepackage{amsthm}
\usepackage{xspace}
\usepackage{enumitem}
\usepackage{subcaption}

\newtheorem{lemma}{Lemma}[section]
\newtheorem{theorem}{Theorem}[section]
\newtheorem{remark}{Remark}[section]
\newtheorem{proposition}{Proposition}[section]

\newcommand{\subgcomp}[1]{{\sc Subgraph complementation to} {#1}}%
\newcommand{\cbdfull}{\subgcomp{$\mathcal{G}_{\Delta \leq k}$}\xspace}%
\newcommand{\cbdk}[1]{{\sc SC to $\Delta \leq {#1}$}\xspace}%
\newcommand{\cbd}{{\sc SC to $\Delta \leq k$}\xspace}%
\newcommand{\Ccbdfull}{\subgcomp{$\mathcal{G}_{\delta \geq k}$}\xspace}%
\newcommand{\Ccbd}{{\sc SC to $\delta \geq k$}\xspace}%

\newcommand{\newproblem}[3]{\bigskip{\centering\fbox{\parbox{0.99\textwidth}{#1\\\textit{Input}: #2\\\textit{Question}: #3}}}\bigskip}

\title{Exploring subgraph complementation to bounded degree graphs}
\author[1]{Ivo Koch}
\author[2]{Nina Pardal}
\author[3]{Vinicius Fernandes dos Santos}

\affiliation[1]{organization={University of General Sarmiento}, country={Argentina}}
\affiliation[2]{organization={University of Huddersfield}, country={United Kingdom}}
\affiliation[3]{organization={Computer Science Department, Federal University of Minas Gerais}, country={Brazil}}

\makeatletter
\def\ps@pprintTitle{%
  \let\@oddhead\@empty
  \let\@evenhead\@empty
  \let\@oddfoot\@empty
  \let\@evenfoot\@oddfoot
}
\makeatother

\begin{document}

\begin{frontmatter}


%
%
\begin{abstract}
Graph modification problems are computational tasks where the goal is to change an input graph $G$ using operations from a fixed set, in order to make the resulting graph satisfy a target property, which usually entails membership to a desired graph class $\mathcal{C}$. Some well-known examples of operations include vertex-deletion, edge-deletion, edge-addition and edge-contraction. In this paper we address an operation known as \emph{subgraph complement}.

Given a graph $G$ and a subset $S$ of its vertices, the subgraph complement of $G \oplus S$ is the graph resulting of complementing the edge set of the subgraph induced by $S$ in $G$.
We say that a graph $H$ is a \emph{subgraph complement} of $G$ if there is an $S$ such that $H$ is isomorphic to $G \oplus S$. 
For a graph class $\mathcal{C}$, the \subgcomp{$\mathcal{C}$} is the problem of deciding, for a given graph $G$, whether $G$ has a subgraph complement in $\mathcal{C}$. 
This problem has been studied and its complexity has been settled for many classes $\mathcal{C}$ such as $\mathcal{H}$-free graphs, for various families $\mathcal{H}$, and for classes of bounded degeneracy. 
In this work, we focus on classes of graphs of minimum/maximum degree upper/lower bounded by some value $k$. In particular, we answer an open question of Antony et al. [Information Processing Letters 188, 106530 (2025)], by showing that \subgcomp{$\mathcal{C}$} is NP-complete when $\mathcal{C}$ is the class of graphs of minimum degree at least $k$, if $k$ is part of the input.
We also show that \subgcomp{$k$-\textsc{regular}} parameterized by $k$ is fixed-parameter tractable. 

\end{abstract}


\end{frontmatter}

\section{Introduction}

Graph modification problems consist of taking an input graph and performing a predefined modification operation to obtain a graph with a desired target property. The goal in these problems is usually to obtain a minimum or minimal set of nodes and/or edges involved in the modification. Graph modification problems
are typically divided into four main classes: completion, deletion, editing (where edges are added, deleted, or both, respectively),  vertex-deletion and vertex-splitting. 
These problems have numerous practical applications across various fields. In computational biology, for example, graph modification problems are used to model relationships between biological entities, such as protein interactions or gene networks. Moreover, these problems are relevant in areas like database theory, where inconsistencies in semi-structured databases may arise from missing or corrupted data, or errors in data processing.

The problem of subgraph complementation was introduced by Kaminski et al. \cite{kaminski2009recent} during their investigations of the clique-width of graphs. 
Subgraph complementation is a particular type of graph modification problem, where the modification operation consists of complementing the edges within a given induced subgraph of the input. Notably, both edge deletion and edge addition can be viewed as special cases of subgraph complementation, since a complementation of the set containing the endpoints of such a (non-)edge would have the same result. 
Perhaps due to the fact that a single operation may affect a large portion of a graph, implying that the structure of the resulting graph may be significantly different from the original one, the study of graph editing by subgraph complementation has been focused mostly on the case when a single complement operation is allowed.
An interesting fact about this operation is that, when applied to a graph class with bounded clique-width, the resulting graphs also have bounded clique-width. This, in turn, allows us to conclude, due to Courcelle's theorem~\cite{Courcelle00lineartime}, that any property expressed in $MSO_1$ within the input class will also be solvable in linear time in the resulting graph. 
Crespelle et al.~\cite{crespelle2023survey} gave a comprehensive survey on the complexity of graph modification problems and results on parameterized algorithms, with a special focus on edge modification problems.


Fomin et al. \cite{fomin2020subgraph} give a number of sufficient conditions for subgraph complementation to a graph class $\mathcal{C}$ to be polynomially solvable, such as $\mathcal{C}$ being a triangle-free class or having bounded degeneracy, as long as $\mathcal{C}$ is recognizable in polynomial time. They also show positive results on graphs of bounded clique-width and on classes defined by some partition matrices and, on the negative side, that the problem is NP-complete when $\mathcal{C}$ is the class of regular graphs. Antony et al.~\cite{antony2022subgraph} investigate the case when $\mathcal{C}$ is the class of $H$-free graphs, where $H$ is a complete graph, a star, a path, or a cycle. 
While the first case is polynomial-time solvable, the remaining three are NP-complete, except for a few cases with small number of vertices. In a later work~\cite{antony2024cutting}, the authors show the NP-completeness of the problem when $\mathcal{C}$ is the class of $H$-free graphs, where $H$ is a tree, with the exception of 41 trees of at most 13 vertices. The authors also find a polynomial-time algorithm for the case in which $H$ is a paw. For the hard cases in~\cite{antony2022subgraph,antony2024cutting}, the authors also prove that they do not admit subexponential-time algorithms (under assumption of the \emph{Exponential Time Hypothesis}). Antony et al.~\cite{antony2023algorithms}  present a polynomial-time algorithm when $\mathcal{C}$ is the class of graphs with minimum degree at least $k$, with $k$ constant, by showing that there exist a kernelization algorithm that returns a kernel linear in $k$. In the same paper, they provide a polynomial-time algorithm for the case when $\mathcal{C}$ is the class of graphs without any induced copies of the star on $k + 1$ vertices (for $k \geq 3$ constant) and the diamond. The authors also ask for the complexity of the following problem: \textit{Given a graph $G$ and an integer $k$, find whether there exists a subgraph complementation of $G$ to a graph with minimum degree at least $k$.}

In this paper, we refine the results concerning minimum and maximum degree of the graph $G$. 
More precisely, we settle the open problem given in \cite{antony2023algorithms} mentioned above by proving the NP-completeness of subgraph complementation to a graph with maximum (resp. minimum) degree at most (resp. at least) $k$. 
In \cite{antony2023algorithms}, the authors' results imply that subgraph complementation to minimum degree at least $k$ is in FPT. 
We draw another connection with the latter work by presenting an FPT algorithm for subgraph complementation to the class of graphs maximum degree at most $k$, parameterized by $k$. We also show that \subgcomp{$k$-\textsc{regular}} parameterized by $k$ is fixed-parameter tractable, which can be interpreted as the intersection of the classes of graphs of maximum and minimum degree $k$.




\section{Preliminaries}

All graphs are finite, undirected and simple. A graph class $\mathcal{C}$ is a set of graphs with a specified property, such as ``being bipartite''. For a graph $G$, we denote by $V(G)$ its vertex set and by $E(G)$ its edge set. We write $\overline{G}$ to denote the complement of the graph $G$. The neighborhood of a vertex $v \in V(G)$ is denoted by $N_G(v)$ and its closed neighborhood $N_G(v) \cup \{v\}$ by $N_G[v]$. Similarly, we denote the neighborhood of a set $S\subseteq V$ by $N_G(S)$, and the closed neighborhood by $N_G\left[ S \right]$. The degree of $v$ is 
denoted by $d(v)$. For any graph $G$, let $\Delta(G)$ be its maximum degree and $\delta(G)$ its minimum degree. A graph is \emph{$d$-degenerate} if every subgraph $H$ of $G$ satisfies $\delta(H) \leq d$. Let $V_{* k}(G) = \{ v \in V(G) \mid d(v) * k \}$, where $* \in \{<,>, =, \neq \}$. 
For all these concepts, we omit the reference to the graph when it is clear from the context. 

As per usual conventions, we denote by $K_n$ the complete graph on $n$ vertices.
The distance between two vertices $u$ and $v$ is the length of a shortest path connecting them. 
The ball of radius $r$ centered in $v$ is the set of all vertices at a distance smaller or equal to $r$ from $v$.

FPT stands for \emph{fixed-parameter tractable}: these are the parameterized problems for which there exists a computable function $f$ such that, given an input $x$ and a parameter $k$, the problem can be solved in $f(k)|x|^{\mathcal{O}(1)}$-time.
The class of parameterized problems that can be solved in $f(k)n^{g(k)}$-time for some computable function $f$ and $g$ is called XP.

We denote by $\oplus$ the subgraph complementation operation, that is, given a graph $G$ and a set of vertices $S \subseteq V(G)$, $G \oplus S$ is the graph obtained from $G$ by complementing all the edges in the subgraph induced by the vertices of $S$, i.e., $V(G \oplus S) = V(G)$ and $E(G \oplus S) = \{uv \in E(G) \mid u \notin S \text{ or } v \notin S\} \cup E(\overline{G[S]})$.
For a fixed graph class $\mathcal{C}$, we define the problem \subgcomp{$\mathcal{C}$} as follows.

\newproblem{\subgcomp{$\mathcal{C}$}}{A graph $G$.}{Is there an $S \subseteq V(G)$ such that $G\oplus S \in \mathcal{C}$?}

In some cases, the definition of a graph class naturally depends on some parameter $k$, like the $k$-colorable graphs. In such cases, it could be convenient to denote it by $\mathcal{C}_k$ and to consider the more general problem where $k$ is part of the input.

\newproblem{\subgcomp{$\mathcal{C}_k$}}{A graph $G$ and $k \in \mathbb{N}$.}{Is there is an $S \subseteq V(G)$ such that $G\oplus S \in \mathcal{C}_k$?}

Let $\pi(G)$ be a graph parameter mapping graphs to naturals, and let $k \in \mathbb{N}$. 
We will denote by $\mathcal{G}_{\pi \leq k}$ the class of graphs $G$ satisfying  $\pi(G) \leq k$. The class $\mathcal{G}_{\pi \geq k}$ is defined analogously. In this paper we are mostly focused on the case when $\pi \in \{\delta, \Delta\}$.

For a graph $G$, it is always possible to create an isolated (universal) vertex through a subgraph complementation, by taking $S=N[v]$ ($S=V(G)\setminus N(v)$) for any $v \in V(G)$, which makes 
\subgcomp{$\mathcal{G}_{\delta \leq k}$} (\subgcomp{$\mathcal{G}_{\Delta \geq k}$}, respectively) trivial.
In the next section we focus on the remaining cases.

\section{Subgraph complementation to bounded degree}

In this section, we will first present a hardness result concerning the complexity of \cbdfull (\cbd), the problem of deciding: given a graph $G$ and a natural number $k$, whether there is a set $S \subseteq V(G)$ such that $\Delta(G\oplus S) \leq k$.
This will then imply hardness results for the problem \Ccbdfull (\Ccbd), which consists of: given a graph $G$ and a natural number $k$, deciding if there exists a subgraph $S$ such that $\delta(G\oplus S) \geq k$.

It is known that \subgcomp{$\mathcal{C}$} is polynomial-time solvable if $\mathcal{C}$ is a class of $d$-degenerate graphs~\cite{fomin2020subgraph}, for some fixed $d$, which in particular includes \subgcomp{$\mathcal{G}_{\Delta \leq k}$} and thus this problem can be solved in polynomial time for any constant $k$. 
The case \subgcomp{$\mathcal{G}_{\delta \geq k}$} has also been shown to be solvable in polynomial time~\cite{antony2023algorithms} for fixed $k$, and in the same paper the case when $k$ is part of the input is stated as an open problem. 
In \cite{fomin2020subgraph} it was also shown that when $\mathcal{C}$ is the class of all regular graphs, the problem is NP-hard. 
It also follows from their construction that \subgcomp{$k$-{\sc regular}} is NP-hard; notice that in this problem we are given both $G$ and $k$ as input, and we want to decide if $G$ can be made $k$-regular.

\begin{figure}
\centering
\includegraphics[width=\textwidth]{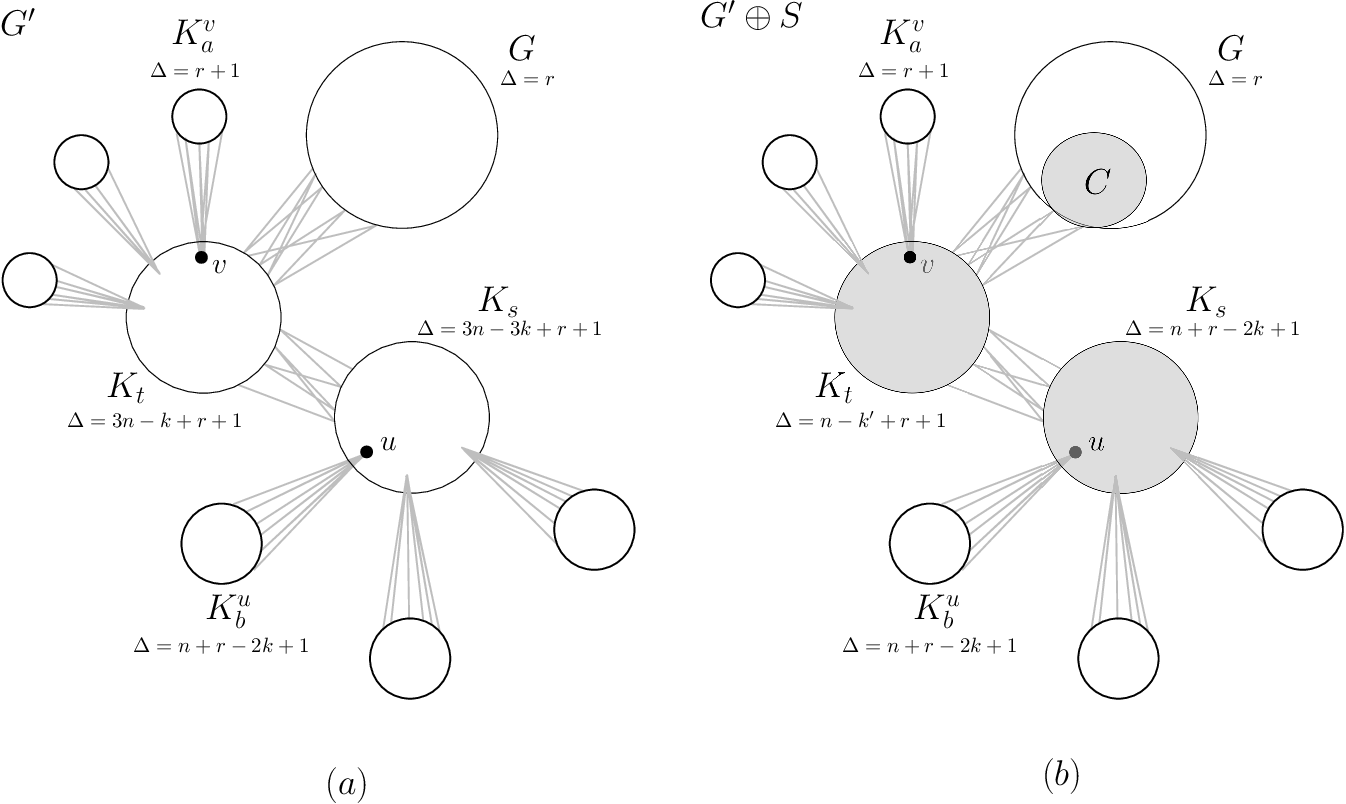}
\caption{Subfigure $(a)$ illustrates the graph reduction used in the proof of Theorem \ref{thm:min_deg_NPc} (the graph $G'$). Except for the circle corresponding to $G$, all circles are complete subgraphs. There are all possible edges between $G$ and $K_t$, and between $K_t$ and $K_s$. Vertices in $K_a^v$ (resp. $K_b^u$) are adjacent to a single vertex $v$ in $K_t$ (resp. $u$ in $K_s$. We state below the name of every subgraph the (maximum) degree $\Delta$ of its vertices. Subfigure $(b)$ shows the vertices of the set $S$ highlighted in gray, as well as the maximum degree of the subgraphs in $G' \oplus S$.} 
\label{fig:reduction}
\end{figure}

\begin{theorem}\label{thm:min_deg_NPc}
     \cbdfull is NP-complete.
\end{theorem}
\begin{proof}
    We show this via a reduction from {\sc Clique on regular graphs (CRG)}. In this problem, given a regular graph $G$ and an integer $k$ asks whether there exists a clique of size $k$ in $G$.  
    Let $(G,k)$ be an instance of {\sc CRG}, where $G$ has $n$ vertices, and $n > k$.  
    Let $r=\Delta(G)$. We may assume $r < n -1$ since otherwise $G$ would be complete and the instance is trivial. We will construct an instance $(G',k')$ of \cbdk{k'} as follows.

    Let $s=n$, $t = n-k+1$, $b = n+r-2k+1$, and $a=r+1$. 
    We start with a copy of $G$ and add the following sets (see Figure \ref{fig:reduction} $(a)$), in this order:
    \begin{enumerate}[label=$\bullet$]
        \item Add (a copy of) $V(K_t)$, and all possible edges between $K_t$ and $G$;
        \item Add (a copy of) $V(K_s)$, and all possible edges between $K_s$ and $K_t$;
        \item For each $v \in V(K_t)$, add a complete graph $K_a^v$ of $a$ vertices, and all possible edges between $K_a^v$ and $v$;
        \item For each $v \in V(K_s)$, add a complete graph $K_b^v$ of $b$ vertices, and all possible edges between $K_b^v$ and $v$.
    \end{enumerate}
    Finally, we set $k'=n+r-k+1$. This completes the definition of the reduction. It is easy to see that the reduction can be done in polynomial time in the size of the input. 
    Suppose $(G,k)$ is a {\sc yes} instance of {\sc CRG}. We will show that $(G',k')$ is a {\sc yes} instance of \cbdk{k'}. Let $C\subseteq V(G)$ be a clique with $|C|=k$. Let $S=C\cup V(K_t)\cup V(K_s)$.
    
    An inspection of the vertex degrees in $G' \oplus S$ (see Figure \ref{fig:reduction} $(b)$ for details) shows that $S$ is a solution for \cbdk{k'}. 
    
    Now for the other direction, let $(G',k')$ be a {\sc yes} instance of \cbdk{k'}, and $W = V(K_t)\cup V(K_s)$.
    First, notice that for any $v \in W$, $d(v) > k'$.
    Hence, for any $S$ with $\Delta(G'\oplus S) \leq k', W \subseteq S$. 
    If $W = S$, then for $v\in K_t$, $d_{G'\oplus S}(v) = n+r+1 > k'$. 
    It follows that every $v\in K_t$ must have additional neighbors in $S$. Assume that those neighbors belong to $K^v_a$. Then the vertices of $K^v_a$ would get a large degree, at least $t - 1 + s$. Since $n > 1$ and $r < n - 1$, this number is greater than $k'$. 
    The same argument can be applied to every $v \in K_s$ with respect to the vertices of $K^v_b$. We conclude that the additional vertices we need for $S$ must stem from $G$, hence we have $S\cap V(G) \neq \emptyset$. Let $C = S\cap V(G)$. We will show that $C$ is a clique.
    For $v\in K_s, d_{G'\oplus S}(v) \geq b+|C| = n+r-2k+1 + |C|$. Since $d_{G'\oplus S}(v) \leq k' = n+r-k+1$, $|C| \leq k$ must hold.
    On the other hand, for $w \in C$, $d_{G'\oplus S}(w) \geq s+r-(|C|-1) = n+r-|C|+1$, where the equality holds only if $w$ is universal to $C$. If $|C| < k$, $d_{G'\oplus S}(w) > k'$. Hence $|C| = k$ and all vertices in $C$ are universal to $C$. Therefore, $C$ is a clique.
\end{proof}

The following remark answers the problem left open in~\cite{antony2023algorithms}.

\begin{remark}\label{rem:max_deg_NPc}
    Using the hardness of \cbd, 
    we can also prove that \Ccbd
    is NP-complete, by taking the instance consisting of $(\overline{G}, n-k-1)$. 
\end{remark}

\section{Natural parameterization and approximation}

We now turn our attention to 
a natural question: \textit{are these problems FPT when parameterized by the (natural) parameter $k$?} 
Recall that \subgcomp{$\mathcal{C}$} is polynomial-time solvable if $\mathcal{C}$ is a class of $d$-degenerate graphs~\cite{fomin2020subgraph} for some $d$. 
Since graphs of bounded maximum degree also have bounded degeneracy, for any fixed $k$ the problem \cbd can be solved in polynomial time. However, the dependence on the degeneracy given by~\cite{fomin2020subgraph} only gives us an XP algorithm. 
On the other hand, the argument of~\cite{antony2023algorithms} gives a linear kernel for the \Ccbd problem, 
implying it is in FPT. Then, a remaining question is to decide if \cbd, 
parameterized by $k$, is 
FPT.
We answer this affirmatively.
In order to show this result, we start by understanding the effect of a subgraph complementation on the degree of a vertex.

\begin{proposition}\label{prop:degree}
    Let $G$ be a graph, and $S \subseteq V(G)$. Then the degree of $v$ in $G \oplus S$ is at least $\max\{|S|-d_G(v)-1,d_G(v) - (|S|-1)\}$ if $v \in S$, and exactly $d_G(v)$ otherwise. 
\end{proposition}
\begin{proof} 
Any vertex $v$ that is not part of $S$ retains its neighborhood intact in $G \oplus S$, that is, $d_{G \oplus S}(v) = d_G(v)$. 
If $v \in S$, it is easy to see that 
$d_{G \oplus S}(v)= |N_G(v) \cup S'| - |N_G(v) \cap S'|$, where $S' = S  \setminus \{v\}$. This sum attains its minimum when $|N_G(v) \cap S'|$ reaches its maximum, i.e. when $N_G(v) \cap S' = N_G(v)$ or $N_G(v) \cap S' = S'$. 
In the first case, we have 
$d_{G \oplus S}(v) = |S'| - |N_G(v)| = (|S| - 1) - d_G(v)$ (since $N_G(v) \subseteq S'$),
and in the second case 
$d_{G \oplus S}(v) = |N_G(v)| - |S'| = d_G(v) - (|S| - 1)$ (since $S' \subseteq N_G(v)$).
\end{proof}

It follows from Proposition~\ref{prop:degree} that, if $\Delta(G \oplus S) \leq k$, then all the vertices in $V_{>k}$ must lie in $S$. 
It is easy to check whether $S = V_{>k}$ gives as a result the desired subgraph complementation, i.e.\, if $\Delta(G \oplus V_{>k}) \leq k$. When this is not the case, we need to check for other solutions. We explore the structure of such an $S$ 
in the following lemma.

\begin{lemma}\label{lemma:smallS}
    Let $G$ be a graph, $k \in \mathbb{N}$, and $S \subseteq V(G)$ such that $\Delta(G\oplus S) \leq k$. If there is a vertex $v \in S$ with $d_G(v) \leq k$,
     then $|S| \leq 2k+1$ and $\Delta(G) \leq 3k$.
\end{lemma}
\begin{proof}
    Let $S \subseteq V(G)$ such that $\Delta(G\oplus S) \leq k$. Recall that, in that case, $V_{>k} \subseteq S$.
    Let $v \in S \setminus V_{>k}$.
    From Proposition~\ref{prop:degree}, we have that $|S|-d_G(v)-1 \leq d_{G\oplus S}(v) \leq \Delta(G\oplus S) \leq k$, which implies $|S| \leq k + d_G(v) + 1$. Since $v \notin V_{>k}$, we know that $d_G(v) \leq k$, and thus it follows that $|S| \leq 2k+1$.

    Furthermore, for any vertex $w \in V(G)$, we also know from Proposition~\ref{prop:degree} that $d_{G\oplus S}(w) \geq d_G(w) - (|S|-1)$, implying $d_G(w) \leq d_{G\oplus S}(w) + |S| - 1 \leq k + (2k+1) - 1 = 3k$ and thus finishing the proof of the lemma.
\end{proof}

    In the decision problem of making the maximum degree bounded by $k$, the natural associated optimization problem would be to minimize the maximum degree.
    As a byproduct, Lemma~\ref{lemma:smallS} provides a straightforward $3$-approximation algorithm for the minimization version of the problem, as stated in the following:


\begin{theorem}\label{thm:3approxMaxDeg}
    There is a 3-approximation polynomial-time algorithm for the problem of finding $S \subseteq V(G)$ that  minimizes $\Delta(G\oplus S)$. 
\end{theorem}
\begin{proof} Consider the following procedure: from $k=0$ to $|V(G)|-1$, we first verify if $\Delta(G\oplus V_{>k}) \leq k$. If true, then we output $k$ as the minimum achievable maximum degree after a subgraph complementation. Otherwise, if $\Delta(G) > 3k$, we conclude by Lemma \ref{lemma:smallS} that $k$ is not achievable and proceed to the next value of $k$. Finally, if $\Delta(G) \leq 3k$, we output $3k$ as the answer. Note that in this case we may assume $S=\emptyset$.
\end{proof}

Although approximation algorithms is not the main focus of this work, we remark that \cite[Lemma 2.1]{antony2023algorithms} implies an approximation for \textsc{Subgraph complementation to}  $\mathcal{G}_{\delta\ge k}$.

\begin{proposition}\label{prop:6approxMinDeg}
    There is a 6-approximation polynomial-time algorithm for the problem of finding $S \subseteq V(G)$ that  maximizes $\delta(G\oplus S)$. 
\end{proposition}

Now, we focus on the parameterized complexity.

\begin{theorem}\label{thm:maxdegFPT}
    There is an FPT algorithm for deciding 
    \textsc{Subgraph complementation to}  $\mathcal{G}_{\Delta\le k}$,
    parameterized by $k$.
\end{theorem}
\begin{proof}We will resort again to  Lemma~\ref{lemma:smallS}.
    We start by checking whether $\Delta(G\oplus V_{>k}) \leq k$ or not. 
    If $V_{>k}$ is not the desired $S$ and $\Delta(G) > 3k$, the algorithm returns ``\textsc{No}'' as an answer. 
    Otherwise, we initialize the set $S$ to $S = V_{>k}$, since $V_{>k} \subseteq S$ for every possible solution. We will branch now on the current vertices of $S$. 
    Since the degree of all vertices that are not in $S$ remains unchanged, any $v$ with $d_{G\oplus S}(v) > k$ must lie in  $S$. For each vertex $v \in S$ such that $d_{G\oplus S}(v) > k$, we know that a vertex $w \in N_G(v)\setminus S$ must be added to $S$ in order to decrease the degree of $v$. 
    We choose one such vertex $v$, and branch on all possible $w$. Since, by Lemma~\ref{lemma:smallS}, a solution exists if it has size $|S|\leq 2k+1$, and we start with $|S| \geq 1$, then the depth of our branching algorithm is at most $2k$. Using the fact that $v$ has at most $\Delta(G) \leq 3k$ neighbors outside $S$ and that $|S|\leq 2k$ in each branching step, we can establish that there are at most $3k$ candidates to be added, thus resulting on a bound of $\mathcal{O}((3k)^{2k})$ subproblems, and therefore yielding an FPT algorithm of time bounded by $\mathcal{O}(9^k k^{2k} |V(G)|^c)$, for some constant $c$.
\end{proof}

The final case we consider is \subgcomp{\sc $k$-regular}, i.e., when one asks for a subgraph complement $H$ such that $\delta(H) = \Delta(H) = k$, which can also be seen as \subgcomp{$\mathcal{G}_{\delta \geq k} \cap \mathcal{G}_{\Delta \leq k}$}. 
It follows from~\cite{fomin2020subgraph} that the problem is 
NP-complete as a consequence of their NP-hardness reduction to \subgcomp{\sc regular}. 
As in the previous case, the problem is solvable in time XP in $k$, since $k$-regular graphs have degeneracy $k$, which leaves the question of whether the problem can be solved in FPT time. We answer this affirmatively. In order to show this, we start by analyzing the structure of the graph induced by the vertices being complemented.

\begin{lemma}\label{lemma:regular-component}
    Let $G$ be a graph, $k \in \mathbb{N}$, let $S \subseteq V(G)$ be such that $G\oplus S$ is $k$-regular and let $C$ be the set of vertices of a connected component of $G[S]$. If $C \subseteq V_{=k}(G)$, then $G[C]$ induces a $\big(\frac{|S|-1}{2}\big)$-regular subgraph.
\end{lemma}

\begin{proof}
    Let $C$ be as in the statement and $v \in C$. 
    Then $|N_{G\oplus S}(v)| = k$, and $N_{G\oplus S}$ can be written as the union of three disjoint sets as
    $$N_{G\oplus S}(v) = (N_G(v)\setminus S) \cup (S\setminus C) \cup (C \setminus (N_G(v) \cup \{v\})).$$
    
    On the other hand, 
    $$N_{G}(v) = (N_G(v)\setminus S) \cup (N_G(v)\cap S)$$
    and $|N_{G}(v)| = k$. 
    Putting these things together, we have
    \begin{equation*}
    \begin{split}
    |N_G(v)\cap S| & = |S\setminus C| + |C \setminus (N_G(v) \cup \{v\})| \\
    & = (|S|-|C|) + (|C|-|N_G(v)\cap C|-1) \\
    & = |S|-|N_G(v)\cap C|-1,
    \end{split}      
    \end{equation*}
    which can be rewritten as 
    $|N_G(v)\cap S| + |N_G(v)\cap C|= |S|-1$.
    As $C \subseteq S$ and since, by the definition of $C$, all neighbors of $v$ in $S$ must belong to $C$, it holds that $|N_G(v)\cap S| = |N_G(v)\cap C|$, and the result follows.
\end{proof}

Notice that, if there is a connected component of $G[S]$ as in the statement of Lemma~\ref{lemma:regular-component}, then that component has more than $|S|/2$ vertices, thus implying that it is unique.
Moreover, it follows similarly as in the discussion preceding Lemma~\ref{lemma:smallS} that, if there is an $S$ such that $G\oplus S$ is $k$-regular, then $V_{\neq k} \subseteq S$.
Hence, from now on we assume that $S$ is the union of two disjoint sets, $S'$ and $C$, where $V_{\neq k } \subseteq S'$, and $C$ is either empty or induces an $\big(\frac{|S|-1}{2}\big)$-regular connected component of $G[S]$. Notice that each connected component of $G[S']$ has at least one vertex in $V_{\neq k}$.

In order to find such an $S$, we proceed as follows: 
we start with $S'=V_{\neq k }$, and proceed by adding vertices adjacent to a previously added vertex. For each candidate $S'$, we look for a suitable (non-empty) $C$. We now show that if such $C$ exists, it can be found in FPT time parameterized by $k$.

\begin{lemma}\label{lemma:regular-component-fpt}
    Let $S' \subseteq V(G)$ be such that $|S'| \leq k$, and each connected component of $G[S']$ has at least one vertex in $V_{\neq k }$. Then we can find, in time FPT in $k$, a set $C$ such that $N(S') \cap C = \emptyset$
    and $G\oplus(S'\cup C)$ is $k$-regular, if it exists.
\end{lemma}
\begin{proof}
    By Lemma~\ref{lemma:smallS}, we may assume that $\Delta(G) \leq 3k$, otherwise $G$ has no $k$-regular subgraph complement. Moreover, since we look for a $C$ such that ${N(S') \cap C = \emptyset}$ and each element of $C$ will be adjacent to each element of $S'$ in  $G\oplus(S'\cup C)$, we know that $|C| \leq k$. 
    Let $v \in V(G) \setminus N(S')$ and let $B_{k-1}(v)$ be the ball of radius $k-1$ around $v$. If $v \in C$, then $C \subseteq B_{k-1}(v)\setminus N(S')$ and since $\Delta(G) \leq 3k$, $|B_{k-1}(v)|$ is bounded by $O((3k)^k)$, which allows us to try all possible subsets of $B_{k-1}(v)$ in FPT time. By doing this for all possible $v \in V(G) \setminus N(S')$ we can find $C$, if it exists, in time $O(|V(G)|^c(3k)^{2k})$ for some constant $c$.
\end{proof}

\begin{theorem}
    \subgcomp{\sc $k$-regular} parameterized by $k$ is in FPT.
\end{theorem}
\begin{proof}

    It follows from Lemma~\ref{lemma:smallS} that, if a solution $S$ transforms $G\oplus S$ to a $k$-regular graph, then $|S| \leq 2k+1$. 
    Starting with $S' = V_{\neq k}$, we test whether $G\oplus S'$ is $k$-regular, which can be done in polynomial time. If not, and if $|S'| \leq k$, by using Lemma~\ref{lemma:regular-component-fpt} we can check whether there exists a $C$ such that $S=S'\cup C$ is a solution in FPT time. Otherwise, for each $v \in N(S')$ we branch on $S'\cup \{v\}$ checking either if $S = S' \cup \{v\}$ is a solution, or whether we can extend it to a solution by using Lemma~\ref{lemma:regular-component-fpt}. Since any solution satisfies $|S| \leq 2k+1$ and $|N(S')| = O(k^2)$, both the breadth and depth of the branch tree are bounded by a function of $k$, and the fact that all operations can be done in time $O(|V(G)|^c(ck)^{ck})$ for some constant $c$ implies that the problem is in FPT. 
\end{proof}

\section{Conclusions and Future work}

In this work, we explored the problem of subgraph complementation to bounded-degree 
graphs, addressing both its classical and parameterized complexity. We demonstrated that the problem of subgraph complementation to graphs with maximum degree at most 
$k$ 
is NP-complete when $k$ is part of the input. This result also implies NP-completeness for the problem of subgraph complementation to graphs with minimum degree at least $k$. 
We also settled the parameterized complexity of these problems for the previously open cases, when parameterized by $k$. Specifically, we showed that the parameterized version of \cbd is FPT. 
Additionally, we presented a $3$-approximation algorithm for the minimization version of this problem, which can be seen as a first step in a new direction for future work. 
Finally, building on previous work, we completed the parameterized complexity analysis for the problem of subgraph complementation to $k$-regular graphs, demonstrating that this problem can be solved in FPT time.

There remain many open questions in the study of modification problems that use complementation as the operation. For instance, the complexity of subgraph complementation to some well-known classes such as chordal or interval graphs remains open. Current ongoing work extends the analysis to other common graph parameters, such as the matching number, dominance number, and diameter. 
Another research avenue would be to restrict the input to specific graph classes, to try and obtain tractable versions of the problems. 

Although in the problems we considered there was a natural parameter related to the vertex degree, in many problems the target class does not have a quantity associated to it. One might consider the size of complemented subgraph $|S|$ as the natural parameter for those classes (and perhaps even for the cases we considered). As far was we know, the only result in this direction is the recent result of Morelle, Sau and Thilikos~\cite{morelle2025graph}, which shows that for a general class of modification problems (which include subgraph complementation), there is an FPT algorithm for the problem of edition to a class $\mathcal{G}$, when $\mathcal{G}$ is a minor-closed graph class. 

The search for approximations algorithms, in the same vein as Theorem~\ref{thm:3approxMaxDeg} and Proposition~\ref{prop:6approxMinDeg}, is also a promising direction that we have only just begun to explore and deserves more attention.
Finally, more general versions of these problems could be explored, focusing on whether a sequence of $k$ subgraph complementations can transform a given graph into one that satisfies the desired property, as usually considered in other graph modification problems.

\section*{Acknowledgements}

\noindent Nina Pardal was partially supported by DFG grant VI 1045/1-1. Vinicius F. dos Santos was partially supported by FAPEMIG and CNPq (Grants 312069/2021-9, 406036/2021-7, and 404479/2023-5).

\bibliographystyle{elsarticle-num}
\bibliography{cas-refs}

\end{document}